\documentclass[12pt]{article}
\usepackage[utf8]{inputenc}
\usepackage{amsfonts,epsfig}
\usepackage[hyphens]{url}
\usepackage{hyperref}
\usepackage{breakurl}
\usepackage[authoryear]{natbib}

\usepackage{geometry} 
\usepackage{sectsty} 
\usepackage[normalem]{ulem} 
\sectionfont{\sffamily\bfseries\upshape\large}
\subsectionfont{\sffamily\bfseries\upshape\normalsize} 
\subsubsectionfont{\sffamily\bfseries\upshape\normalsize} 
\makeatletter
\usepackage{graphicx}
\usepackage{wrapfig}
 \usepackage{graphicx,fullpage,amsmath,multicol,multirow,amssymb,amsbsy,pifont}
 \usepackage{setspace}
 \usepackage{epsfig}
 \usepackage{amsmath, amsthm, amssymb, bm}
 \usepackage{color}
\usepackage{graphics}
\usepackage{booktabs}
\usepackage{makecell}
\usepackage{caption,array}

\renewcommand\@seccntformat[1]{\csname the#1\endcsname.\quad}

\numberwithin{figure}{section}
\numberwithin{table}{section}
\numberwithin{equation}{subsection} 
\makeatletter
\@addtoreset{equation}{section}
\makeatother
\makeatletter
\def\@maketitle{%
 \begin{center}%
 \let \footnote \thanks
  {\Large \@title \par}%
  {\normalsize
   \begin{tabular}[t]{c}%
    \@author
   \end{tabular}\par}%
  {\small \@date}%
 \end{center}%
}
\makeatother

\title{\bf Fully Synthetic Data for Complex Surveys}

\author{Shirley Mathur\thanks{Department of Statistics, B-313 Padelford Hall, University of Washington, Seattle, WA 98195-4322} \and  Yajuan Si\thanks{Survey Research Center, Institute for Social Research, University of Michigan, Rm 4014, 426 Thompson St., Ann Arbor, MI 48104} \and Jerome P. Reiter\thanks{Department of Statistical Science, 214a Old Chemistry Building, Duke University, Durham, NC 27708-0251 }}
\date{}

\begin{document}
\maketitle

\begin{abstract}
When seeking to release public use files for confidential data, statistical agencies can generate fully synthetic data. We propose an approach for making fully synthetic data from surveys collected with complex sampling designs. Our approach adheres to the general strategy proposed by \citet{rubin:1993}. Specifically, we generate pseudo-populations by applying the weighted finite population Bayesian bootstrap to account for survey weights, take simple random samples from those pseudo-populations, estimate synthesis models using these simple random samples, and release simulated data drawn from the models as public use files. To facilitate variance estimation, we use the framework of multiple imputation with two data generation strategies. In the first, we generate multiple data sets from each simple random sample. In the second, we generate a single synthetic data set from each simple random sample. We present multiple imputation combining rules for each setting. We illustrate the repeated sampling properties of the combining rules via simulation studies, including comparisons with synthetic data generation based on pseudo-likelihood methods. We apply the proposed methods to a subset of data from the American Community Survey.


\end{abstract}

\section{Introduction}\label{intro}

Many national statistics agencies, survey organizations, and researchers---henceforth all called agencies---disseminate microdata, i.e., data on individual units, to the public. Wide dissemination of microdata greatly benefits society, enabling broad subsets of the research community to access and analyze the collected data \citep{reiterisr}. Often, however, agencies cannot release microdata as collected, because doing so could reveal survey respondents' identities or values of sensitive attributes, thereby failing to satisfy ethical or legal requirements to protect data subjects' confidentiality \citep{reiter:raghu:07}. 

To manage these risks, several agencies have implemented or are considering synthetic data approaches, as first proposed by \citet{rubin:1993}. In this approach, the agency (i) randomly and independently samples units from the sampling frame to comprise each synthetic data set, (ii) imputes the unknown data values for units in the synthetic samples using models fit with the original survey data, and (iii) releases multiple versions of these data sets to the public. These are called fully synthetic data sets \citep{drechslerbook, raghu:arisa}. Releasing fully synthetic data can preserve confidentiality, since identification of units and their sensitive data can be difficult when the released data are not actual, collected values \citep{2stagesyn}. Methods for inferences from these multiply-imputed data files have been developed for a variety of statistical inference tasks~\citep{raghu:rubin:2001, reiter:2002,reiter:2002a,reitermulti, reit:drech:census, sireiter11}.  

While prominent applications of fully synthetic data exist for censuses or administrative data \citep[e.g., ][]{lbdisr}, 
many research data sets are based on surveys collected with sampling designs that use unequal probabilities of selection. Previous research on multiple imputation for missing data suggests that imputation models should account for the survey design features, such as stratification, clustering, and survey weights \citep{reitragukin}. Similarly, when using multiple imputation for synthetic data, the models also should account for the survey design \citep{mitra:reiter:2006, Weightsforprivacy:Fienberg10, SynInfor:Kim2021}. The key challenge is properly incorporating weights in the synthesis models, which relates to the long-standing debate about the role of survey weights in model-based inferences \citep{pfeffermann93,Pfeffermann11, little04-model}. 

Researchers have proposed a variety of approaches for generating fully synthetic data in complex surveys. The suggestion in early work \citep{rubin:1993, raghu:rubin:2001, reiter:2002} was to take a Bayesian finite population inference approach, in which the agency (i) builds predictive models for the survey variables conditional on design features like stratum/cluster indicators or size measures, which are assumed known by the agency for every unit in the population, (ii) imputes the missing survey variables for the nonsampled units in the population, and (iii) takes a simple random sample from the completed population to release as one synthetic data set. A related approach uses the weighted finite population Bayesian bootstrap (WFPBB) \citep{fpbb14}, in which the agency generates completed populations by replicating individuals from the confidential data in proportion to their survey weights and then releases the completed populations, forgoing the step of simple random sampling. More recently, it has been suggested to build synthetic data models that account for the sampling design directly, so that they estimate the joint distribution of the population data. For example, the agency can use a pseudo-likelihood approach \citep{pfeffermann93, Savitsky16}, in which each individual's contribution to the likelihood function of a synthesis model is raised to a power that is a function of the survey weights \citep{SynInfor:Kim2021}. Departing from the proposal of \citet{rubin:1993}, a completely different approach is to create and attach new weights to synthetic data records simulated from models that are agnostic to the survey weights \citep{unecesynthetic}. Here, the goal is to allow users to use weighted estimates that scale up to the finite population. The new weights can be created by treating the survey weights as a variable in the synthesis, so that the agency specifies a predictive model for the weights. The simulated weights may be adjusted by raking or calibration before inclusion in the released file.   
 
Each of these methods has its potential drawbacks. The Bayesian finite population inference approach, while theoretically principled, requires completing full populations, which can be cumbersome, and the availability of design variables for all records in the population, which may not be the case in some surveys. The WFPBB releases (multiple copies of) individuals' genuine data records, which creates obvious disclosure risks. Pseudo-likelihood approaches may not estimate sampling variability correctly~\citep{Williams:Savitsky:ISR21}, and it is not clear how easily they can be implemented with machine learning synthesizers like classification and regression trees \citep{reitercart}, which are commonly used in practical synthetic data projects \citep{raab2016practical}. With synthesized weights, secondary analysts are expected to use the simulated weights to approximate design-based inference. This approximation does not have a theoretical basis; as such, it is unclear whether the synthetic weights approach facilitates accurate inferences in general.

In this article, we propose an approach to generate fully synthetic data from complex samples in the spirit of the original proposal of \cite{rubin:1993}, i.e., the agency releases simple random samples that do not require users to perform survey-weighted analyses with the synthetic data. To do so, we build on the WFPBB approach of \citet{fpbb14} by first creating pseudo-populations that account for the survey weights.  We then take simple random samples (SRSs) from each pseudo-population, estimate synthesis models from each SRS, and generate draws from these models to create multiply-imputed, fully synthetic public use files. The latter step provides confidentiality protection, as the agency is not releasing genuine records. We consider two processes for the last step of generating the synthetic data.  In {\em Synrep-R}, we generate multiple synthetic data sets from each SRS.  In {\em SynRep-1}, we generate one synthetic data set from each SRS. {\em SynRep-R} releases more data sets than {\em SynRep-1}, which can result in reduced variances.  However, the additional data sets can increase the overhead for the agency and secondary analysts, and they provide additional information for adversaries seeking disclosures. For both approaches, we derive multiple imputation combining rules that enable the estimation of variances. Using simulation studies, we illustrate the repeated sampling performances of the combining rules and compare them to fully synthetic data generated while disregarding the sampling design entirely. We also compare them against approaches that use synthesis models estimated with weighted pseudo-likelihoods \citep{SynInfor:Kim2021}. Finally, we illustrate the proposed methods using a subset of the American Community Survey (ACS) data. Code for the simulation studies and the ACS illustration is available at \href{https://github.com/yajuansi-sophie/SynRep}{(GitHub address to be included in the final paper)}.

The remainder of the article is organized as follows. Section~\ref{method} describes the two synthetic data generation processes in detail and presents the new combining rules. Section~\ref{simulation} presents the simulation studies. Section~\ref{application} presents the illustration with the ACS data.  Section~\ref{discussion} suggests topics for future research. 

\section{Proposed Methods for Generating Fully Synthetic Survey Data}
\label{method}

\begin{figure}
\begin{tabular}{c}
\includegraphics[width=0.95\textwidth]{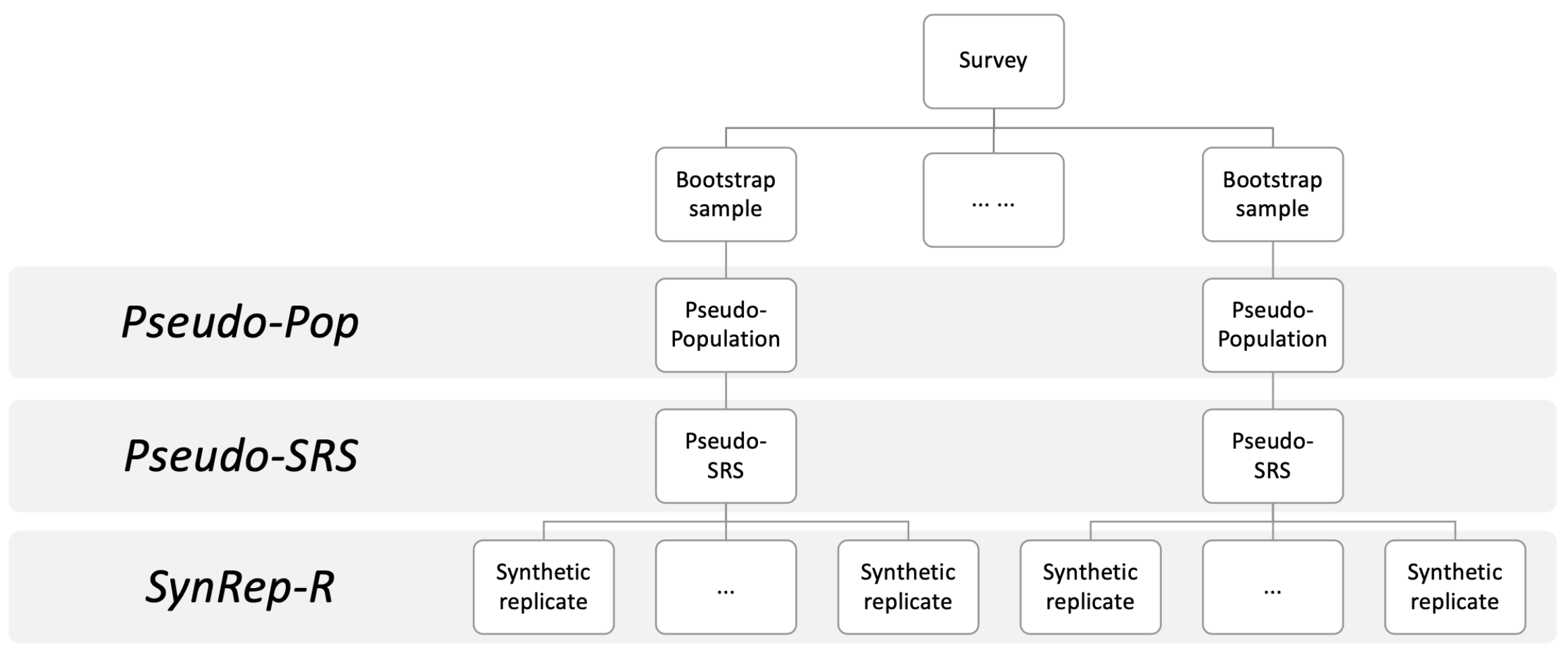}
\end{tabular}
\caption{Process for generating synthetic data with multiple data sets per simple random sample (SRS), which we call {\em SynRep-R}.}
\label{2stage}
\end{figure}

\begin{figure}
\begin{tabular}{c}
\includegraphics[width=0.95\textwidth]{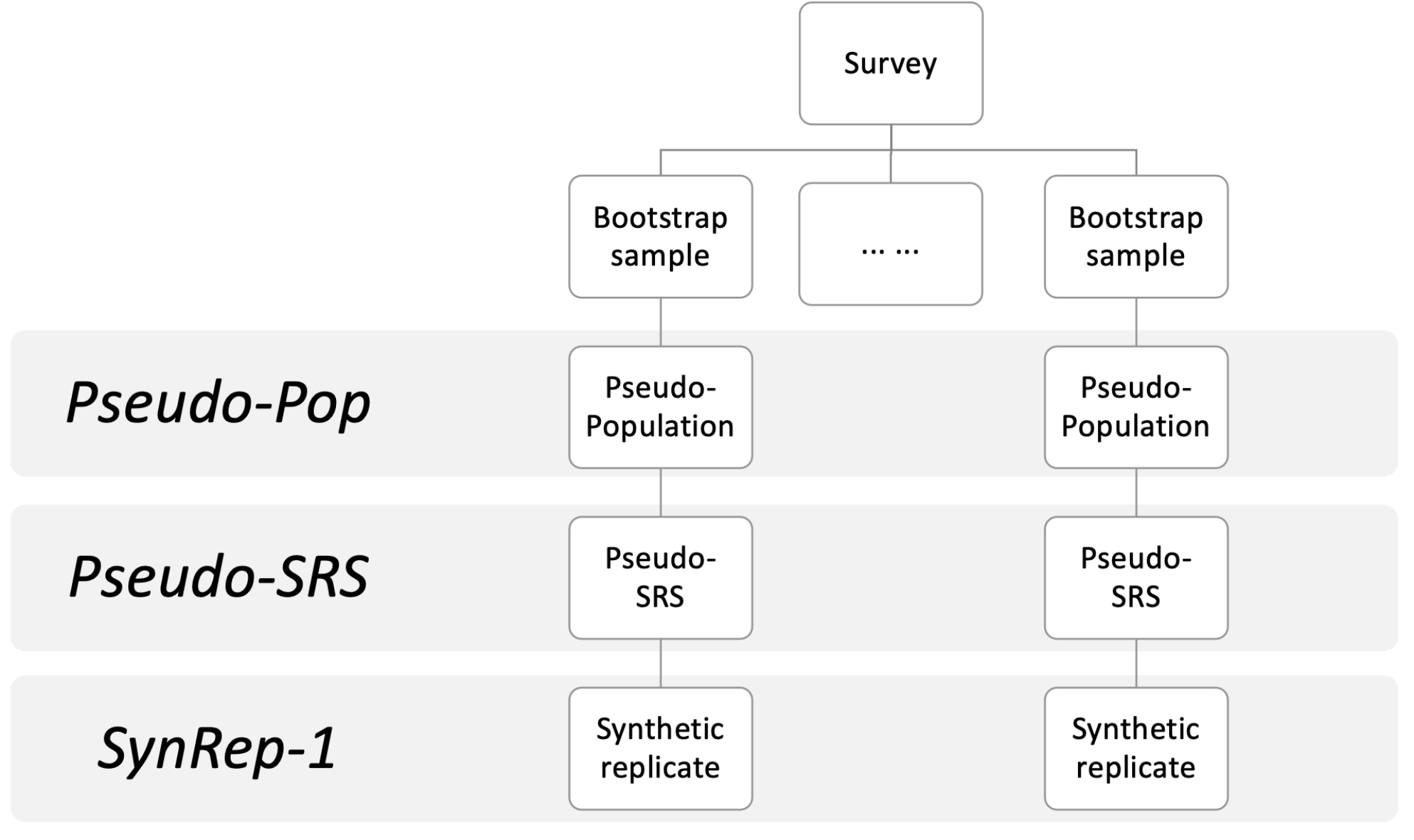}
\end{tabular}
\caption{Process for generating synthetic data with one data set per simple random sample (SRS), which we call {\em SynRep-1}.}
\label{1stage}
\end{figure}

Let $\mathcal{D}$ be a probability sample of size $n$ randomly drawn from a finite population comprising $N$ units. For $i=1, \dots, N$, let $\pi_i$ be the selection probability for unit $i$, and let $w_i=1/\pi_i$ be the unit's survey weight.  Here, we are agnostic as to whether $w_i$ is potentially adjusted, e.g.,  for calibration or nonresponse, although in our simulation studies we use pure design weights.  For $i=1, \dots, N$, let $Y_i$ be the $p \times 1$ vector of survey variables.  Hence, $\mathcal{D} = \{(w_i,Y_i): i = 1, \dots, n\}$. For simplicity of exposition, we suppose that $p=1$, so that $Y_i$ is a scalar.  {\em SynRep-R} and {\em SynRep-1}, and their corresponding inferential methods, can be used with multivariate survey data as well.  

In Section \ref{method:generate}, we describe the processes of generating synthetic data. In Section \ref{method:inference}, we describe the inferential methods. As mentioned in Section \ref{intro} and following the proposal in \citet{rubin:1993}, we take as a goal allowing secondary users to analyze the released data sets as if they were simple random samples from the population. 

\subsection{Data generation process}\label{method:generate}

Figure \ref{2stage} and Figure \ref{1stage} display the processes of generating synthetic data for {\em SynRep-R} and {\em SynRep-1}, respectively. We now describe these steps in detail.

In either process, the first step is to generate pseudo-populations using the WFPBB  \citep{fpbb14}. The WFPBB generates pseudo-populations by ``undoing'' the complex sampling design and accounting for the sampling weights. The idea is to draw from the posterior predictive distribution of non-observed data ($Y_{nob}$) given the observed data ($Y_{obs}$) and the survey weights, i.e., drawing from $P(Y_{nob}\mid  Y_{obs}, w_1, \dots, w_n)$. This distribution supposes that the population is comprised of the unique values of $Y_i \in \mathcal{D}$, and that the corresponding counts for each value in the population follow a multinomial distribution. With a non-informative Dirichlet prior distribution on the multinomial probabilities, the P\'{o}lya distribution can be used to draw the predictive samples in place of the Dirichlet-multinomial distribution.  

With this in mind, the process of generating the synthetic data is described below.
\begin{enumerate}
    \item \textbf{Resample via Bayesian bootstrap}: To inject sufficient  sampling variability,
    using the data from the ``parent'' sample $\mathcal{D}$, we generate \textit{M} samples, $(\mathcal{S}^{(1)}, \ldots, \mathcal{S}^{(M)}),$ each of size $n$ using independent Bayesian bootstraps \citep{rubin:bb:81}. For each $\mathcal{S}^{(m)}$ and for $i=1, \dots, n$, let $w_i^{(m)}=c w_i r^{(m)}_{i}$, 
    where $r^{(m)}_{i}$ is the number of times that element $i$ from $\mathcal{D}$ appears in $\mathcal{S}^{(m)}$.  The $c$ is a normalizing constant to ensure that the new weights sum to the population size $N$. Thus, in each $\mathcal{S}^{(m)}$, for $i=1, \dots, n$, we create  $w^{(m)}_i=(Nw_i r^{(m)}_{i})/(\sum_k w_k r^{(m)}_k)$. 

    \item \textbf{Use the WFPBB to make pseudo-populations}:  For each $\mathcal{S}^{(m)}$, we construct an initial P\'{o}lya urn  using the set of $\{Y_i, w^{(m)}_i\}$. We then draw $N-n$ units using probabilities $(p^{(m)}_1, \dots, p^{(m)}_n)$ determined from 
    \begin{equation}
     p^{(m)}_i=  \frac{w^{(m)}_i - 1 + l^{(m)}_{i, k-1}(N-n)/n}{N-n + (k-1)(N-n)/n},
        \label{eqpolya}
     \end{equation}
for the $k$th draw, $k \in \{1, \ldots, N-n\}$, where $l^{(m)}_{i, k-1}$ is the number of bootstrap selections of $Y_i$ among the elements present in the urn at the $k-1$ draw. The $N-n$ draws combined with the data in $\mathcal{S}^{(m)}$ comprise one pseudo-population, $\mathcal{P}^{(m)}$.  We repeat this  for $m=1, \ldots, M$ to create $\mathcal{P}_{pseudo}= \{\mathcal{P}^{(m)}: m=1, \dots, M\}.$  When $N$ is very large, we can save memory and computational costs by creating a pseudo-population that is large enough to be practically the same for inference as a population of size $N$, which we operationalize by generating $50n$ rather than $N-n$ records.
    \item \textbf{Draw SRS from each pseudo-population}: For $m=1, \dots, M$, take a simple random sample $\mathcal{D}^{(m)}$ of size $n$ from $\mathcal{P}^{(m)}$. Let $\mathcal{D}_{srs} = \{\mathcal{D}^{(m)}: m=1, \dots, M\}.$
    \item \textbf{Generate synthetic data replicates:} For $m=1, \dots, M$, estimate a synthesis model using  $\mathcal{D}^{(m)}$, and 
 draw from the predictive distributions to form synthetic data replicates using either Step 4a or Step 4b.  
    \begin{itemize}
        \item[4a.] \textbf{\em SynRep-R:} For $m=1, \dots, M$, draw $R>1$ synthetic replicates $\mathcal{D}^{(m,r)}_{syn}$ of size $n$, where $r =1, \ldots, R$, using each $\mathcal{D}^{(m)}.$  We release $\mathcal{D}_{syn} = \{\mathcal{D}^{(m,r)}_{syn}: m=1, \dots, M; r=1, \dots, R\}$ including  indicators of which $m$ each $\mathcal{D}^{(m,r)}_{syn}$ belongs to.  
        \item[4b.] \textbf{\em SynRep-1:} For $m=1, \dots, M$, draw one synthetic data sample $\mathcal{D}^{(m)}_{syn}$ of size $n$ from each $\mathcal{D}^{(m)}$.  Release $\mathcal{D}_{syn} = \{\mathcal{D}^{(m)}_{syn}: m=1, \dots, M\}.$
    \end{itemize}
The synthesis model for each $\mathcal{D}^{(m)}$ can utilize plug-in values of model parameters, e.g., their maximum likelihood estimates. It is not necessary to use posterior distributions at this stage of the process \citep{kinneyreiterjos}.

\end{enumerate}

As these two processes for generating synthetic data differ from those of \citet{raghu:rubin:2001}, as well as from other synthetic data scenarios such as those of  \citet{reiterpartsyn} and \citet{reitermi}, we require new methods for inferences, to which we now turn.

\subsection{Inferences for {\em SynRep-R} and {\em SynRep-1}}
\label{method:inference}


To derive the inferential methods, we follow the general strategy of multiple imputation \citep{rubin:1987} and use a Bayesian inference approach. For any population quantity $Q$, such as the population mean $Q \equiv\bar{Y}$, we seek the posterior distribution $P(Q \mid \mathcal{D}_{syn})$. Following \cite{raghu:rubin:2001}, we compute the following integral based upon each level of the data synthesis process from Figure \ref{2stage} or Figure \ref{1stage}.
\begin{align}
    P(Q \mid \mathcal{D}_{syn}) 
    &= \int \int \int P( Q  \mid  \mathcal{D}_{syn}, \mathcal{D}_{srs}, \mathcal{P}_{pseudo}, \mathcal{D} ) P( \mathcal{D}  \mid  \mathcal{D}_{syn}, \mathcal{D}_{srs}, \mathcal{P}_{pseudo} ) \notag\\
    &\qquad\qquad P( \mathcal{P}_{pseudo}  \mid  \mathcal{D}_{syn}, \mathcal{D}_{srs}) P( \mathcal{D}_{srs}  \mid  \mathcal{D}_{syn} ) d \mathcal{D} d \mathcal{P}_{pseudo} d \mathcal{D}_{srs}.\label{eq:initial}
\end{align}

When we condition on $\mathcal{D}$, the values of $(\mathcal{D}_{syn}, \mathcal{D}_{srs}, \mathcal{P}_{pseudo})$ do not provide any additional information about $Q$.    
Thus, we can simplify $P( Q  \mid  \mathcal{D}_{syn}, \mathcal{D}_{srs}, \mathcal{P}_{pseudo}, \mathcal{D}) = P(Q  \mid  \mathcal{D})$. When we condition on $\mathcal{P}_{pseudo}$, the values of $(\mathcal{D}_{rep}, \mathcal{D}_{syn})$ provide no additional information about $\mathcal{D}$.  Thus, we simplify $P( \mathcal{D}  \mid  \mathcal{D}_{syn}, \mathcal{D}_{srs}, \mathcal{P}_{pseudo}) = P(\mathcal{D}  \mid  \mathcal{P}_{pseudo})$. When we condition on  $\mathcal{D}_{srs}$, the value of $\mathcal{D}_{syn}$ provides no information about $\mathcal{P}_{pseudo}$. Hence, $P(\mathcal{P}_{pseudo}  \mid  \mathcal{D}_{syn}, \mathcal{D}_{srs} ) = P(\mathcal{P}_{pseudo}  \mid  \mathcal{D}_{srs})$.
With some re-arrangement to aid interpretation, we re-express \eqref{eq:initial} as 
\begin{align}
\label{eq:full_integral}
    P(Q \mid \mathcal{D}_{syn}) = \int & \left[  \int \left[ \int  P(Q  \mid  \mathcal{D}) P(\mathcal{D}  \mid  \mathcal{P}_{pseudo}) d \mathcal{D} \right] 
     P(\mathcal{P}_{pseudo}  \mid  \mathcal{D}_{srs}) d \mathcal{P}_{pseudo} \right] \notag\\
   & P(\mathcal{D}_{srs}  \mid  \mathcal{D}_{syn}) d \mathcal{D}_{srs}.
\end{align}


We begin with  $P(Q \mid \mathcal{P}_{pseudo}) = \int P( Q  \mid  \mathcal{D} ) P( \mathcal{D}  \mid  \mathcal{P}_{pseudo} ) d \mathcal{D}.$  
We assume that, for large $M$, this is approximately a normal distribution. This should be reasonable in large samples, which are typical in settings where agencies want to release public use data. We only require the posterior distribution of $Q$ to be normal, not the distribution of the survey variables themselves; indeed, the underlying data can be categorical.  We note that the inferential methods are not intended for quantities like medians or other quantiles; inferential methods for such quantities is a topic for additional research.

We only require means and variances to characterize normal sampling distributions. Thus, we focus on estimating the distributions of the first two moments.  For $m=1, \dots, M$, let $Q^{(m)}$ be the computed value of $Q$ if we had access to $\mathcal{P}^{(m)}$. \cite{rubin:1987} shows that
\begin{align}
\label{P-syn}
    (Q \mid \mathcal{P}_{pseudo}) \sim t_{M-1}\left(\bar{Q}, \left(1 + M^{-1}\right)B\right),
\end{align}
where $\bar{Q} = \sum_{m}Q^{(m)}/M$ and $B = \sum_{m} \left(Q^{(m)} - \bar{Q} \right)^2/(M-1)$.
Here $t_{\nu}(\mu, \sigma^2)$ denotes a $t$-distribution with $\nu$ degrees of freedom, location $\mu$, and variance $\sigma^2$. In the derivations, for convenience we approximate the  $t$-distribution in \eqref{P-syn} as a normal distribution, which should be reasonable for somewhat large $M$.


We next turn to $P(\mathcal{P}_{pseudo}  \mid  \mathcal{D}_{srs})$.  Here, we only need $P(\bar{Q}, B  \mid  \mathcal{D}_{srs})$. For $m = 1, \dots, M$, let $q^{(m)}$ be the estimate of $Q^{(m)}$ and  $v^{(m)}$ be the estimate of the sampling variance associated with $q^{(m)}$; we could compute these if we had access to $\mathcal{D}^{(m)}$. We assume that $\{q^{(m)}, v^{(m)}: m=1, \dots, M\}$  are valid in the following sense.
\begin{enumerate}
    \item[1)] For each $m$, $q^{(m)}$ is approximately unbiased for $Q^{(m)}$ and asymptotically normally distributed, with respect to repeated sampling from the pseudo-population $\mathcal{P}^{(m)}$ with sampling variance $V^{(m)}$.  That is, we have $(q^{(m)} \mid \mathcal{P}^{(m)}) \sim N(Q^{(m)}, V^{(m)})$.
    \item[2)] The sampling variance estimate $v^{(m)}$ is approximately unbiased for $V^{(m)}$, and the sampling variability in $v^{(m)}$ is negligible.  That is, $(v^{(m)} \mid \mathcal{P}^{(m)}) \approx V^{(m)}$.
    \item[3)] The variation in $V^{(m)}$ across the $M$ pseudo-populations is negligible; that is, $V^{(m)} \approx V \approx \bar{v}$, where $\bar{v} = \sum_m v^{(m)}/M$.
\end{enumerate} 

Using standard Bayesian arguments based on these sampling distributions, it follows that
\begin{align}
(Q^{(m)} \mid q^{(m)}, \bar{v}) \, &\overset{ind}{\sim} \, N(q^{(m)}, \bar{v})\\
(\bar{Q} \mid \bar{q}, \bar{v}) \, &\sim \, N(\bar{q}, \bar{v}/M),
\end{align}
where $\bar{q}=\sum_m q^{(m)}/M$. 

To obtain the distribution of $(Q \mid \mathcal{D}_{srs})$, we integrate the distribution in $\eqref{P-syn}$, which we approximate as a normal distribution, with respect to the distributions of $\bar{Q}$ and $B$. We only need the first two moments since the resulting distribution is a  normal distribution.  We have 
\begin{equation}
E(Q\mid \mathcal{D}_{srs}) = E(E(Q \mid \bar{Q})\mid \mathcal{D}_{srs})=E(\bar{Q}\mid \mathcal{D}_{srs}) = \bar{q}.\label{eq:srsexpected}
\end{equation}
We also have 
\begin{align}
\label{v-d-syn}
\nonumber Var(Q \mid \mathcal{D}_{srs})&=E(Var(Q \mid \mathcal{P}_{pseudo})\mid \mathcal{D}_{srs}) + Var(E(Q \mid \mathcal{P}_{pseudo})\mid \mathcal{D}_{srs})\\
&= (1+M^{-1}) E(B\mid \mathcal{D}_{srs}) + \bar{v}/M.
\end{align}
This is the variance estimator in \citet{raghu:rubin:2001}, which analysts would use if the agency releases $\mathcal{D}_{srs}$ as the public use files.  However, since we take an additional step of replacing each $\mathcal{D}^{(m)}$ with simulated values, we need to average over the distributions of $(\bar{q}, \bar{v}, B)$.  The result depends on whether we use {\em SynRep-R} or {\em SynRep-1}, as we now describe.


\subsubsection{Derivation with  {\em SynRep-R}}\label{sec:synrepr}

For each $\mathcal{D}^{(m,r)}_{syn}$, let $q^{(m,r)}_{syn}$ be the point estimate of $Q$, and let $v^{(m,r)}_{syn}$ be the estimate of the variance  associated with $q^{(m,r)}_{syn}$.  The analyst computes $q^{(m,r)}_{syn}$ and $v^{(m,r)}_{syn}$ acting as if $\mathcal{D}^{(m,r)}_{syn}$ is the collected data obtained via a simple random sample of size $n$ taken from the population. The analyst needs to compute the following quantities.
\begin{align}
    \bar{q}^{(m)}_{syn} &= \sum_{r=1}^R q^{(m,r)}_{syn}/R \label{eq:q_m_bar_def} \\
    \bar{q}_{syn} &= \sum_{m=1}^M \bar{q}^{(m)}_{syn}/
    M\label{eq:q_bar_def} \\
    b_{syn} &= \sum_{m=1}^M (\bar{q}^{(m)}_{syn} - \bar{q}_{syn})^2/(M-1) \label{eq:b_def}\\
    w^{(m)}_{syn} &= \sum_{r=1}^R (q^{(m,r)}_{syn} - \bar{q}^{(m)}_{syn})^2/(R-1) \label{eq:w_m_def} \\
    \bar{w}_{syn} &= \sum_{m=1}^M w^{(m)}_{syn}/M \label{eq:w_bar_def}\\
     \bar{v}_{syn} &= \sum_{m=1}^M \sum_{r=1}^R v^{(m,r)}_{syn}/MR. \label{eq:v_bar_def}
\end{align}

We now complete the derivation of the posterior distribution for $(Q \mid \mathcal{D}_{syn})$ in the {\em SynRep-R} approach. To do so, we assume large-sample normal approximations for the sampling distributions of the point estimates. Specifically, for all $(m,r)$, we assume that 
\begin{equation}
q^{(m,r)}_{syn} \, \sim \, N(q^{(m)}, W^{(m)}),\label{samplingw}
\end{equation} 
where $W^{(m)}$ is the sampling variance for $q^{(m,r)}_{syn}$ over draws of synthetic data from $\mathcal{D}^{(m)}$. The normality should be reasonable when $n$ is large. Assuming diffuse prior distributions and conditioning on $W^{(m)}$, we have 
\begin{align}
(q^{(m)} \mid \mathcal{D}^{(m,1)}_{syn}, \dots,  \mathcal{D}^{(m,R)}_{syn}, W^{(m)}) \, & \sim \, N(\bar{q}^{(m)}_{syn}, W^{(m)}/R)\\ 
(\bar{q} \mid \mathcal{D}_{syn}, \bar{W}) \, & \sim \, N(\bar{q}_{syn}, \bar{W}/MR),\label{qbardist}
\end{align}
where $\bar{W} = \sum_m W^{(m)}/M.$ 

Having now determined distributions for the point estimators, we put everything together for the posterior distribution of $Q$. Since all the components are normal distributions, $P(Q \mid \mathcal{D}_{syn})$ is a normal distribution. Thus, for the expectation, we use \eqref{eq:srsexpected} and \eqref{qbardist} to obtain 
\begin{equation}
E(Q\mid \mathcal{D}_{syn}) = E(E(Q \mid \mathcal{D}_{srs})\mid \mathcal{D}_{syn})=E(\bar{q}\mid \mathcal{D}_{syn}) = \bar{q}_{syn}.
\end{equation}


For the variance, we first write the variance in terms of $(B, \bar{v}, \bar{W})$ and then plug in point estimates of these terms.  To emphasize the use of $(B, \bar{v}, \bar{W})$, we write
\begin{align}
Var(Q\mid \mathcal{D}_{syn}, B, \bar{v}_M, \bar{W})& \nonumber= E((1+M^{-1})B + \bar{v}/M) \mid \mathcal{D}_{syn}, B, \bar{v}, \bar{W}) + Var(\bar{q} \mid \mathcal{D}_{syn}, B, \bar{v}, \bar{W})\\
& = (1+M^{-1})B + \bar{v}/M + \bar{W}/MR.\label{varfinal}
\end{align}



We now define the estimates for $(B, \bar{v}, \bar{W})$, which we plug into \eqref{varfinal}.  For $\bar{v}$, we assume that $\bar{v}_{syn} \approx \bar{v}$. 
This assumption follows from the rationale in \cite{raghu:rubin:2001}, who argue this is the case when the synthetic data are generated from the same underlying distribution as the data used to fit the models.

For $\bar{W}$, we note that  \eqref{samplingw} implies that, for $m=1, \dots, M$,  
\begin{equation}
\frac{(R-1)w_{syn}^{(m)}}{W^{(m)}} \,\sim \, \chi^2_{R-1}.\label{chisqw}
\end{equation}
We further assume that each $W^{(m)} \approx \bar{W}$. This assumption is in line with a similar assumption provided in \cite{reitermi} regarding the variability of posterior variances. Essentially, as stated in \cite{reitermi}, this assumption stems from the observation that variability amongst posterior variances is generally smaller in magnitude than variability in posterior expectations.  With this assumption and utilizing \eqref{chisqw}, we have 
\begin{align}
    \sum_{m=1}^M \frac{(R-1)w^{(m)}_{syn}}{\bar{W}} &\sim \chi^2_{M(R-1)}.
\end{align}
Thus, we have 
\begin{equation}
    \text{E}\left(\sum_{m=1}^M \frac{(R-1)w^{(m)}_{syn}}{\bar{W}}\right) = M(R-1). 
\end{equation}    
Utilizing a methods of moments approach to approximate $\bar{W}$, we obtain $\bar{W} \approx \bar{w}_{syn}$.

For approximating $B$, we note that the sampling distribution of a randomly generated $\bar{q}^{(m)}_{syn}$ over all steps in the data generation process is $N(Q, B + \bar{v} + \bar{W}/R)$. Using this fact, we have 
\begin{align}
 \frac{\sum_{m=1}^M(\bar{q}^{(m)}_{syn} - \bar{q}_{syn})^2}{B + \bar{v} + \bar{W}/R} \sim \chi^2_{M-1}, 
\end{align}
so that
\begin{align}
    \text{E}\left(\frac{\sum_{m=1}^M(\bar{q}^{(m)}_{syn} - \bar{q}_{syn})^2}{B + \bar{v} + \bar{W}/R}\right) &= M-1 . 
\end{align}
Using a method of moments approach and the definition of $b_{syn}$ in (\ref{eq:b_def}), and the plug-in estimate $\bar{w}_{syn}$ for $\bar{W}$, we have $b_{syn} \approx  B + \bar{v}_{syn} + \bar{w}_{syn}/R$, so that $B \approx b_{syn} - \bar{v}_{syn} - \bar{w}_{syn}/R.$

Putting all together, we can approximate $\text{Var}(Q \mid \mathcal{D}_{syn})$ with the estimate $T_r$, where 
\begin{align}
\label{eq:Tr_def}
    T_r & = \left(1+M^{-1} \right)\left( b_{syn} - \bar{v}_{syn} - \bar{w}_{syn}/R\right)+ \bar{v}_{syn}/M + \bar{w}_{syn}/MR  \notag\\
    &= \left(1+M^{-1} \right)b_{syn} - \bar{v}_{syn} - \bar{w}_{syn}/R.
\end{align}

We compute approximate 95\% intervals for $Q$ as $\bar{q}_{syn} \pm t_{0.975, M-1}\sqrt{T_r}$.  The $t$-distribution is a simple approximation based on the degrees of freedom in \eqref{P-syn}. As with the variance estimator in \citet{raghu:rubin:2001}, the estimate $T_r$ can be negative, particularly for small $M$.  As an {\em ad hoc} adjustment when $T_r<0$, we recommend replacing $B$ with $\bar{v}$ in \eqref{varfinal} and using $T^*_r=\left(1+2/M  \right){\bar{v}_{syn}} + \bar{w}_{syn}/MR$.

\subsubsection{Derivation with {\em SynRep-1}}\label{sec:synrep1}

With large $M$ and $R$, {\em SynRep-R} results in many synthetic data sets, which may be undesirable from the perspective of the agency and secondary data analysts. Instead, agencies may want to use {\em SynRep-1}.  To obtain inferences for $Q$ in this setting, we leverage the methodology of \citet{raab2016practical}, who observed that when the source data come from a simple random sample, as is the case for each $\mathcal{D}^{(m)}$, we can obtain valid variance estimates with single implicates with adjustments of the combining rules. We now describe this derivation.
 
 
 For $m=1, \dots, M$, let $q^{(m)}_{syn}$ be the point estimate of $Q$ computed using $\mathcal{D}_{syn}^{(m)}$, and let $v^{(m)}_{syn}$ be the estimated variance associated with $q^{(m)}_{syn}$.  The analyst computes each $(q^{(m)}_{syn}, v^{(m)}_{syn})$ by acting is if $\mathcal{D}_{syn}^{(m)}$ is a SRS of size $n$ from the population.  We require the following quantities for inferences.  To economize on notation, we re-use some of the notation introduced in Section \ref{sec:synrepr}.   
\begin{align}
    \bar{q}_{syn} &= \sum_{m=1}^M q_{syn}^{(m)}/M \label{eq1:q_bar_def} \\
     b_{syn} &= \sum_{m=1}^M ({q}_{syn}^{(m)} - \bar{q}_{syn})^2/(M-1) \label{eq1:b_def} \\
    \bar{v}_{syn} &= \sum_{m=1}^M v_{syn}^{(m)}/M. \label{eq1:v_bar_def}
\end{align}
The pairs of equations \eqref{eq1:q_bar_def} and \eqref{eq:q_bar_def}, \eqref{eq1:b_def} and \eqref{eq:b_def}, and \eqref{eq1:v_bar_def} and \eqref{eq:v_bar_def} can be viewed as equivalent when $R=1$.


To complete the derivation for {\em SynRep-1}, we follow the logic in \citet{raab2016practical} and assume that 
$q^{(m)}_{syn} \sim N(q^{(m)}, V^{(m)})$.
Assuming $V^{(m)} \approx \bar{v}$ for all $m$,  we have 
\begin{align}
    (q^{(m)} \mid \mathcal{D}_{syn}^{(m)}, \bar{v}) & \,\sim \, N(q_{syn}^{(m)}, \bar{v})\\
(\bar{q} \mid \mathcal{D}_{syn}, \bar{v}) & \,\sim \, N(\bar{q}_{syn}, \bar{v}/M). \label{eq:qbarsynrep1}
    \end{align}
We note, however, that one should not assume that $B \approx \bar{v}$ as well. As $\mathcal{D}$ is a complex sample, it yields sampling variances that could differ from the simple random sampling variances associated with  $\mathcal{D}_{srs}.$ 

 Since all the components are approximately normal distributions, $P(Q \mid \mathcal{D}_{syn})$ also is approximately a normal distribution. For its expectation, we use \eqref{eq:srsexpected} and \eqref{eq:qbarsynrep1} to obtain 
\begin{equation}
E(Q\mid \mathcal{D}_{syn}) = E(E(Q \mid \mathcal{D}_{srs})\mid \mathcal{D}_{syn})=E(\bar{q}\mid \mathcal{D}_{syn}) = \bar{q}_{syn}.
\end{equation}

For its variance, as with {\em SynRep-R}, we write the variance in terms of $(B, \bar{v})$ and then plug in point estimates of these terms. We have
\begin{align}
Var(Q\mid \mathcal{D}_{syn}, B, \bar{v}) &\nonumber = E((1+M^{-1})B + \bar{v}/M) \mid \mathcal{D}_{syn}, B, \bar{v})  + Var(\bar{q} \mid \mathcal{D}_{syn}, B, \bar{v})\\
&= (1+ M^{-1})B + \bar{v}/M + \bar{v}/M 
 = (1+ M^{-1})B + 2\bar{v}/M. \label{varsynrep1}
\end{align}

We now define the estimates for $(B, \bar{v})$ to plug into \eqref{varsynrep1}.  For $\bar{v}$, we use $\bar{v}_{syn}$ defined in \eqref{eq1:v_bar_def}.  This should be reasonable since we are replacing the entire set of each $\mathcal{D}^{(m)}$ with synthetic values. 
To approximate $B$, we note that the sampling distribution of a randomly generated $q^{(m)}_{syn}$ over all steps in the data generation process is $N(Q, B + 2\bar{v})$. Using this fact, we have 
\begin{align}
 \frac{\sum_{m=1}^M({q}^{(m)}_{syn} - \bar{q}_{syn})^2}{B + 2\bar{v}} \sim \chi^2_{M-1}, 
\end{align}
so that
\begin{align}
    \text{E}\left(\frac{\sum_{m=1}^M({q}^{(m)}_{syn} - \bar{q}_{syn})^2}{B + 2\bar{v}}\right) &= M-1 . 
\end{align}
Using a method of moments approach and the definition of $b_{syn}$ in (\ref{eq1:b_def}), we have $b_{syn} \approx  B + 2\bar{v}_{syn}$, so that $B \approx b_{syn} - 2\bar{v}_{syn}.$

Thus, we can approximate $Var(Q \mid \mathcal{D}_{syn})$ with the estimate $T_m$, where  
\begin{align}
\label{eq1:Tm_def}
T_m = \left(1+M^{-1} \right)b_{syn} - 2{\bar{v}_{syn}}.
\end{align} 
We compute approximate 95\% intervals for $Q$ as $\bar{q}_{syn} \pm t_{0.975, M-1}\sqrt{T_m}.$  When $T_m<0$, as an {\em ad hoc} variance estimate we replace $B$ by $\bar{v}$ in \eqref{varsynrep1} and use $T^*_m=\left(1+3/M \right){\bar{v}_{syn}}$.

\section{Simulation Studies}
\label{simulation}

In this section, we present simulation studies to illustrate the repeated sampling properties of the inferential methods in Section \ref{method:inference} for various finite population quantities.  

\subsection{Simulation design} \label{sim:design}

We construct a finite population based on data from the Public Use Microdata Sample of the 2021 American Community Survey \citep{acs2021}.  This file comprises $3,252,599$ individuals, which we treat as a population of size $N$. The file also has person-level weights (named ``PWGTP'' in the data file). We do not treat these as survey weights, per se; rather, we treat them as size variables $x_i$, where $i=1, \dots, N$, for use in probability proportional to size (PPS) sampling. We also use these constructed size measures to generate two survey variables, $(y_{i1}, y_{i2})$, where $i=1, \dots, N$. Specifically, we let each $y_{i1}$ be a binary variable sampled from a Bernoulli distribution with probability $\textrm{Pr}(y_{i1}=1) =\exp(-7+2\log x_i)/\left(1+\exp(-7+2\log x_i)\right)$. We let each $y_{i2}$ be a continuous variable sampled from a normal distribution with mean $20+50y_{i1}$ and standard deviation $50$. We estimate the finite population proportion $\bar{Y}_1 = \sum_{i=1}^Ny_{i1}/N \approx .765$; the finite population mean $\bar{Y}_2 = \sum_{i=1}^Ny_{i2}/N \approx 58.2$; and, the finite population regression coefficient of $Y_1$ in the linear regression of $Y_2$ on $Y_1$, which is $\beta \approx 50$.

From this population, we sample  $\mathcal{D}$ using a PPS sample of size $n=500$ survey units, setting $\pi_i = n x_i/\sum_{i=1}^Nx_i$ and using the function ``ppss'' in the $R$ package ``pps''~\citep{R:pps}. Under this PPS sampling design, we expect that unweighted inferences using $\mathcal{D}$ should be badly biased for $(\bar{Y}_1, \bar{Y}_2)$ but perhaps not so for $\beta$. We repeat the sampling process to create 1000 independent realizations of $\mathcal{D}$.

For each $\mathcal{D}$, we implement  {\em SynRep-R} and {\em SynRep-1} with various $(M, R)$. Specifically, we examine $(M=4, R=5)$, $(M=10, R=5)$, $(M=50, R=5)$, $(M=10, R=10)$, $(M=10, R=25)$, and $(M=10, R=50)$.  The choice of $R$ only affects {\em SynRep-R}.  We implement the WFPBB using the ``polyapost''  package in $R$~\citep{polyapost}, creating pseudo-populations $(\mathcal{P}^{(1)}, \dots, \mathcal{P}^{(M)})$ each comprising 25,000 individuals. 
From each $\mathcal{P}^{(m)}$ where $m=1, \dots, M$, we take a simple random sample of size $n$ to make a corresponding  $\mathcal{D}^{(m)}$. To make each synthetic data replicate stemming from each $\mathcal{D}^{(m)}$, we sample $n$ synthetic values for $Y_1$ using a Bernoulli distribution with probability set to the empirical proportion of $Y_1$ in $\mathcal{D}^{(m)}$. We sample the corresponding synthetic values of $Y_2$ from normal distributions with means equal to the predicted values from the regression of $Y_2$ on $Y_1$, computed using the synthetic values of $Y_1$ and the unbiased estimates of the coefficients computed with $\mathcal{D}^{(m)}$, and variance equal to the unbiased estimate of the regression variance computed with $\mathcal{D}^{(m)}$.

To assist in evaluating the repeated sampling performances of {\em SynRep-1} and {\em SynRep-R}, we also use results computed with  $\mathcal{P}_{pseudo}$ and $\mathcal{D}_{srs}$. Specifically, in each of the 1000 simulation runs, we define {\em Pseudo-Pop} as the procedure that uses a point estimator of $\bar{Q}$ and variance estimator of $(1+1/M)B$ computed with the WFPBB-generated pseudo-populations $(\mathcal{P}^{(1)}, \dots, \mathcal{P}^{(M)})$.  We define {\em Pseudo-SRS} as the procedure that uses a point estimator of $\bar{q}$ and variance estimator of \cite{raghu:rubin:2001} computed with $(\mathcal{D}^{(1)}, \dots, \mathcal{D}^{(M)})$. As a comparison against what happens if we disregard the sampling design entirely, we define {\em SRSsyn} as the procedure that generates synthetic data by using (i) the unweighted sample proportion for $Y_1$ as the Bernoulli probability to generate $n$  synthetic values of $Y_1$ and (ii) the unweighted estimates of parameters in the regression of $Y_2$ on $Y_1$ as the parameters of the normal distribution to generate the corresponding $n$ synthetic values of $Y_2$.

We also evaluate the repeated sampling performances of pseudo-likelihood approaches to making fully synthetic data. For each synthesis model, i.e., the Bernoulli and linear regression models, we start with a likelihood function defined as the product of the contributions from each individual in $\mathcal{D}$.  We create the pseudo-likelihood by raising each individual's contribution to a power defined by the individual's survey weight. We use these weighted pseudo-likelihoods to estimate synthesis model parameters. We implement this approach using the software $Stan$~\citep{stan-software:2024}, which can generate posterior samples of model parameters based on user-specified likelihood functions.
We run $Stan$ to create four chains of 4,000 iterations and discard the first 2,000 iterations as burn-in. We randomly sample one of the resulting draws and use its parameter values in the Bernoulli and linear regression models to generate the synthetic data. We repeat this process $M$ times and apply the inference rules in \cite{raghu:rubin:2001}.  We call this method {\em Wtreg}. We note that that \cite{SynInfor:Kim2021}  use the variance estimator \eqref{v-d-syn} from  \cite{raghu:rubin:2001} with $\bar{v}=0$.  \cite{SynInfor:Kim2021} release synthetic populations (where $\bar{v}=0$) rather than synthetic samples (where $\bar{v}>0$).

We also consider a modification of {\em Wtreg} to address potential underestimation of variability in the parameter draws.  We call this method {\em Wtreg-Boot}. First, we take a bootstrap sample of size $n$ from $\mathcal{D}$.  We construct the pseudo-likelihood functions using the bootstrapped data and the survey weight for each resampled individual. Using this pseudo-likelihood function, we then generate and analyze synthetic data following the steps described for {\em Wtreg}.

Finally, we define {\em Direct} as using the unweighted sample mean and standard deviation from $\mathcal{D}$, i.e.,  ignoring the survey weights, and {\em HT} as using the \citet{ht52} estimator and its estimated variance using $\mathcal{D}$.  We use these latter two procedures to assess the importance of accounting for the sampling design in inferences with $\mathcal{D}$.

Let superscript $s$ index the results from simulation run $s$, where $s=1, \dots, 1000$. For any estimator $\hat{q}$ for any of the methods we examine, we compute the percent bias, $100\sum_{s=1}^{1000}(\hat{q}^s - Q)/(1000Q)$. We compute the proportion of the 1000 95\% confidence intervals based on $\hat{q}$ and its corresponding variance estimate that cover $Q$. We also compute the ratio of the empirical variance of the 1000 values of $\hat{q}$ to the empirical variance of the 1000 values of the {\em HT} point estimator. To investigate the accuracy of variance estimators, for each method we compute the ratio of the average of the 1000 variance estimates over its corresponding empirical variance. Finally, to examine the stability of the variance estimator for each method, we compute the standard deviation of the 1000 variance estimates. We present results for the first four quantities in the main text and for the last quantity in the Appendix.

\subsection{Results}

\begin{figure}[htp]
\caption{Repeated sampling properties of {\em SynRep-1} and {\em SynRep-R} for $\bar{Y}_1$ under different numbers of synthetic samples ($M$) and replicates ($R$) under a PPS design.}
\label{pps-x}
\begin{tabular}{c}
\includegraphics[width=0.9\textwidth]{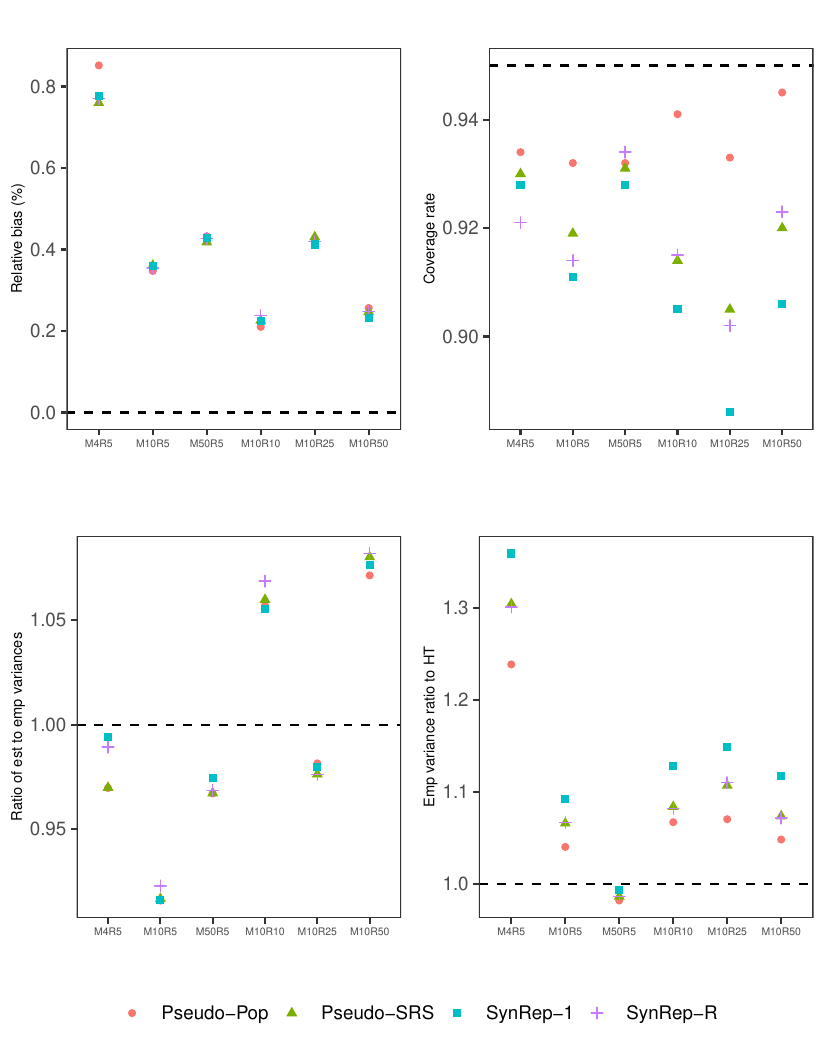}
\end{tabular}
\end{figure}

\begin{figure}[htp]
\caption{Repeated sampling properties of {\em SynRep-1} and {\em SynRep-R} for $\bar{Y}_2$ under different numbers of synthetic samples ($M$) and replicates ($R$) under a PPS design.}
\label{pps-y}
\begin{tabular}{c}
\includegraphics[width=0.9\textwidth]{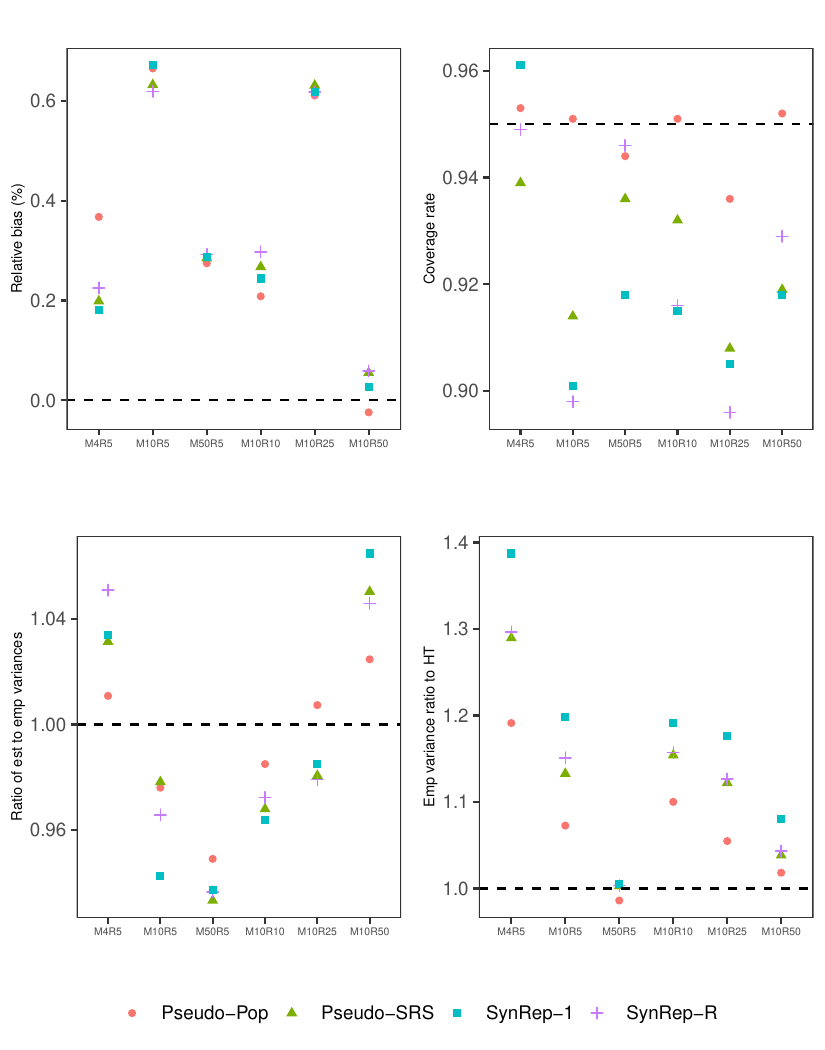}
\end{tabular}
\end{figure}

\begin{figure}[htp]
\caption{Repeated sampling properties of {\em SynRep-1} and {\em SynRep-R} for $\beta$ under different numbers of synthetic samples ($M$) and replicates ($R$) under a PPS design.}
\label{pps-beta}
\begin{tabular}{c}
\includegraphics[width=0.9\textwidth]{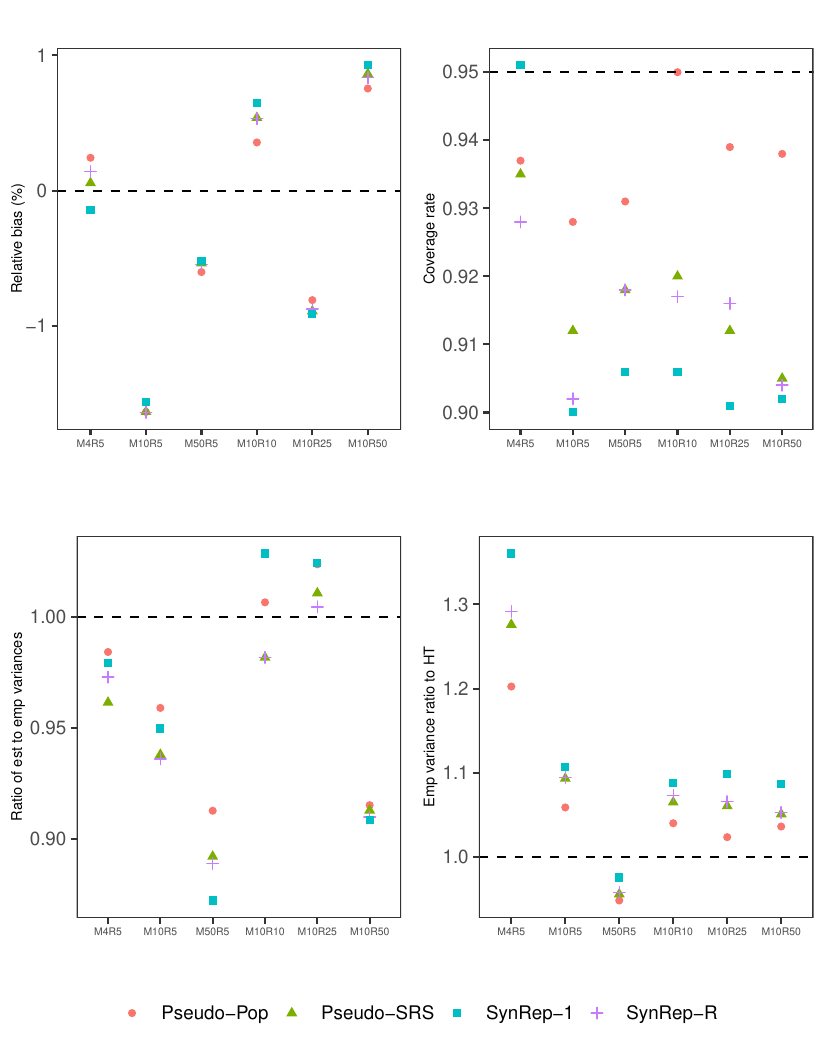}
\end{tabular}
\end{figure}

We first investigate the properties of {\em SynRep-R} and {\em SynRep-1} for the various settings of $(M,R)$. Figure \ref{pps-x}, Figure~\ref{pps-y}, and Figure~\ref{pps-beta} display results for $\bar{Y}_1$, $\bar{Y}_2$, and $\beta$, respectively, for these two methods as well as for {\em Pseudo-Pop} and {\em Pseudo-SRS}. All four methods offer approximately unbiased point estimates of the three finite population quantities, with simulated percent biases generally around 1\% or lower. 
These small biases originate primarily from the step of completing populations, as the biases in {\em Pseudo-Pop} are close to the biases in the other three methods.  As expected, compared to the variance for {\em HT}, the simulated variances are increasingly inflated as $M$ decreases. Holding $M=10$ constant,  decreasing $R$ tends to increase the simulated variances, although the effects are less pronounced than those from decreasing $M$. The variability in {\em SynRep-1} results with fixed $M$ reflects Monte Carlo error. Taken together, these results suggest it is preferable to increase $M$ rather than $R$ when keeping $MR$ constant. For example, when we compare {\em SynRep-R} with $(M=10, R=5)$ to {\em SynRep-1} with $M=50$, the latter tends to result in smaller empirical variance with closer-to-nominal coverage rates. Similar benefits appear when comparing {\em SynRep-R} with $(M=10, R=25)$ to {\em SynRep-R} with $(M=50, R=5)$. This finding accords with results from \citet{mandr}, who considered a similar trade-off for nested multiple imputation for partially synthetic and missing data. We note that using larger values of $M$ also offers smaller variability in the estimated variances, as shown in the Appendix.  

By comparing the ratios of the empirical variances to the variances for {\em HT}, we can see the effect on efficiency of the steps in the synthesis process. The variances generally increase as we go from {\em Pseudo-Pop} to {\em Pseudo-SRS} to {\em SynRep-R} or {\em SynRep-1}; that is, they increase as we add more steps that involve randomness. The variances for {\em SynRep-R} generally are slightly smaller than those for {\em SynRep-1}, reflecting the benefit for efficiency of the additional information from $MR$ rather than $M$ synthetic data sets.  We note that the variance inflation from using synthetic data procedures versus {\em HT} largely disappears when $M=50$.

Across all four synthetic data methods, the average variance estimates are reasonably similar to the empirical variances.  Disparities from ratios of one apparently stem, once again, mainly from the step of completing the populations. The confidence interval coverage rates range from a low of 88\% to a high of 96\%, with most slightly below nominal. Coverage rates for {\em SynRep-R} and {\em SynRep-1} tend to be highest when $M=50$, further reflecting the benefits of using a larger $M$. For $M \geq 10$, the coverage rates for {\em SynRep-R} tend to be higher than those for {\em SynRep-1}, although the difference is typically only a point or two.

\begin{table}[tp]
\centering
\caption{Proportion of negative variance estimates in the PPS simulation studies. When $M=50$, all variance estimates are positive.}
\label{neg-T}
\begin{tabular}{llccc}
  \hline
 $(M,R)$ & Method & $\bar{Y}_1$ & $\bar{Y}_2$ & $\beta$ \\ 
  \hline
M4R5 & Pseudo-SRS & 0.09 & 0.15 & 0.13 \\ 
M4R5 & SynRep-R & 0.11 & 0.17 & 0.13 \\ 
M4R5 & SynRep-1 & 0.17 & 0.26 & 0.22 \\ 
M10R5 & Pseudo-SRS & 0.01 & 0.02 & 0.02 \\ 
M10R5 & SynRep-R & 0.01 & 0.03 & 0.02 \\ 
M10R5 & SynRep-1 & 0.04 & 0.09 & 0.07 \\ 
   \hline
\end{tabular}
\end{table}

The combining rules in~\eqref{eq1:Tm_def} and \eqref{eq:Tr_def} do result in negative variance estimates, as evident in Table~\ref{neg-T}. In the simulations, we use $T_r^*$ and $T_m^*$ to make confidence intervals when needed.  As $M$ increases, the number of negative variance estimates decreases.  In fact, when $M=50$, all of the variance estimates are positive, offering additional support for making $M$ large.  The estimates of $b_{syn}$ become less variable as $M$ increases, which helps avoid the negative variances. Negative variance rates tend to be lower for {\em SynRep-R} than for {\em SynRep-1}, reflecting the benefits of increased datasets to estimate variance parameters. Although not shown in Table~\ref{neg-T},  the negative variance rates when $M=10$ do not change much as we increase $R \geq 5$.  We note that the negative variance rates for {\em SynReg-R} are similar to those for {\em Pseudo-SRS}. Evidently, when $MR$ is large, the information available in $\mathcal{D}_{syn}$ to estimate $b_{syn}$ is on par with the information available in $\mathcal{D}_{srs}$.  

\begin{figure}[htp]
\caption{Repeated sampling properties of different quantities and procedures with $M=10$ synthetic samples and $R=10$ replicates under a PPS design. }
\label{pps-m10r10}
\begin{tabular}{c}
\includegraphics[width=0.95\textwidth]{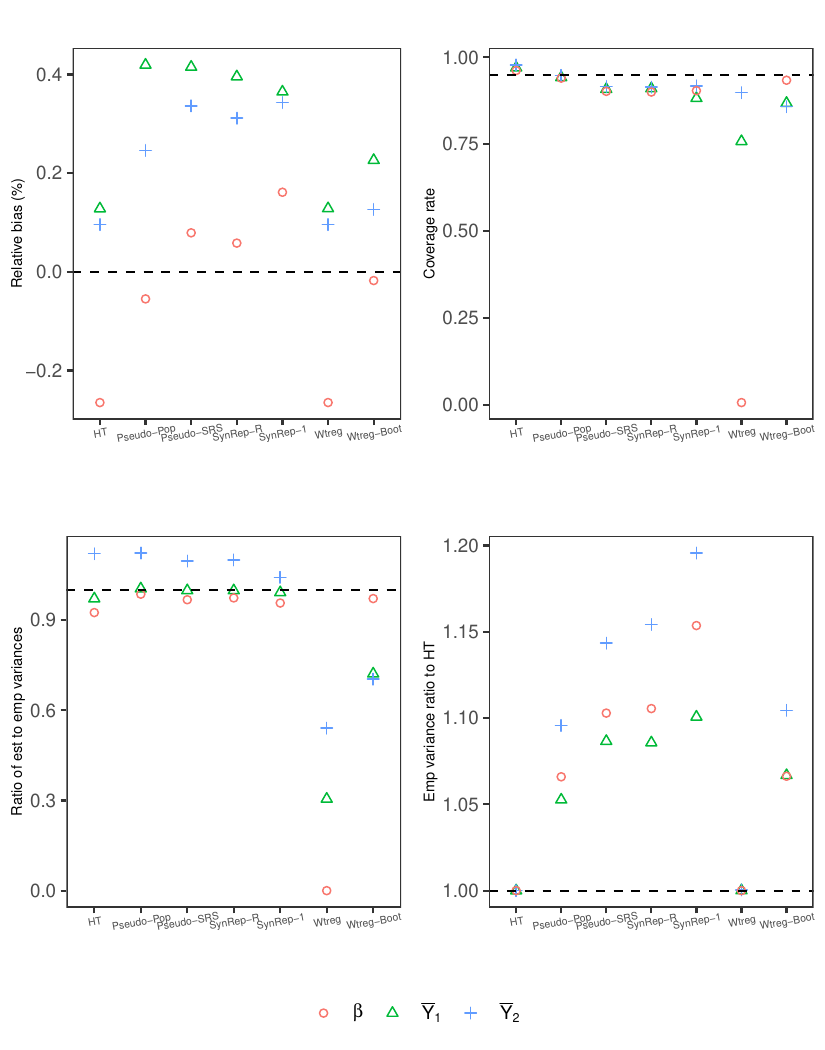}
\end{tabular}
\end{figure}

We next turn to compare {\em SynRep-R} and {\em SynRep-1} with other approaches, particularly {\em Wtreg}, {\em Wtreg-Boot}, and {\em SRSsyn}. Here, we set $M=10$ and, where relevant, $R=10$, and draw 500 repeated samples.  Figure~\ref{pps-m10r10} summarizes the repeated sampling performances of the methods that account for survey weights. For all these methods, the point estimators have simulated percent biases that typically are negligible. For {\em SynRep-R} and {\em SynRep-1}, the average variance estimates are close to their corresponding empirical variances, and the coverage rates are close to nominal.  For {\em Wtreg} and {\em Wtreg-Boot}, the variance estimators can underestimate the corresponding empirical variances severely, especially for $\bar{Y}_1$ and $\bar{Y}_2$, resulting in confidence interval coverage rates that can be substantially lower than the nominal 95\% level. The bootstrap step in {\em Wtreg-Boot} results in more reliable variance estimates compared to {\em Wtreg}, but {\em Wtreg-Boot} is not as well calibrated as {\em SynRep-R} and {\em SynRep-1}, which have closer to nominal coverage rates. As expected, {\em HT} results in accurate estimates with near nominal coverage rates. We note that Figure~\ref{pps-m10r10} does not display results for {\em Direct} and {\em SRSsyn} because they perform poorly for $\bar{Y}_1$ and $\bar{Y}_2$. For these two methods, the simulated biases for $\bar{Y}_1$  and for $\bar{Y}_2$ are around $16$\% and $11$\%, respectively, with coverage rates near 0 and near 30\%, respectively. These results emphasize the importance of accounting for informative designs when generating fully synthetic data that can be analyzed as simple random samples.

Overall, the simulation studies suggest that {\em SynRep-R} and {\em SynRep-1} can provide approximately valid inferences, and they are superior inferentially to fully synthetic data that ignore the complex design. The Appendix also includes results of simulation studies where we sample $\mathcal{D}$ via simple random samples. These confirm that the combining rules offer reasonable performance even without unequal probabilities of selection.

\section{Illustration with ACS Data}
\label{application}

We illustrate {\em SynRep-R} and {\em SynRep-1} by letting $\mathcal{D}$ be a subset of data from the 2021 ACS Public Use Microdata Sample for $n=84,128$ individuals from the state of Michigan. The variables for our illustration include each participant's person-level weight, age, and total income.  To mimic the variables in the simulations, we create a binary indicator $Y_1$ from age that equals one when someone is at least 65 years old; we refer to this indicator as senior status.  For purposes of synthesis, we transform income by taking its cubic root. The synthesis models are then a Bernoulli distribution for $Y_1$ and a linear regression of the cubic root of total income on $Y_1$. After synthesizing values of the cubic root of income, we raise them to the third power to get incomes on the original scale.  We implement each method following the procedures from Section~\ref{simulation}. For {\em SynRep-R} and {\em SynRep-1}, we set $M=10$ and $R=10$.

As population quantities, we estimate the population proportion of senior status individuals $\bar{Y}_1$, the population mean of the income values, $\bar{Y}_2$, and the coefficient $\beta$ of $Y_1$ in the linear regression model of the cubic-root transformed income on senior status.

\begin{figure}[htp]
\caption{Point estimates and 95\% confidence intervals for $\bar{Y}_1$, $\bar{Y}_2$, and $\beta$ in the ACS data illustration. Results based on $M=10$ synthetic samples and $R=10$ replicates.}
\label{acs-m10r10}
\begin{tabular}{c}
\includegraphics[width=0.95\textwidth]{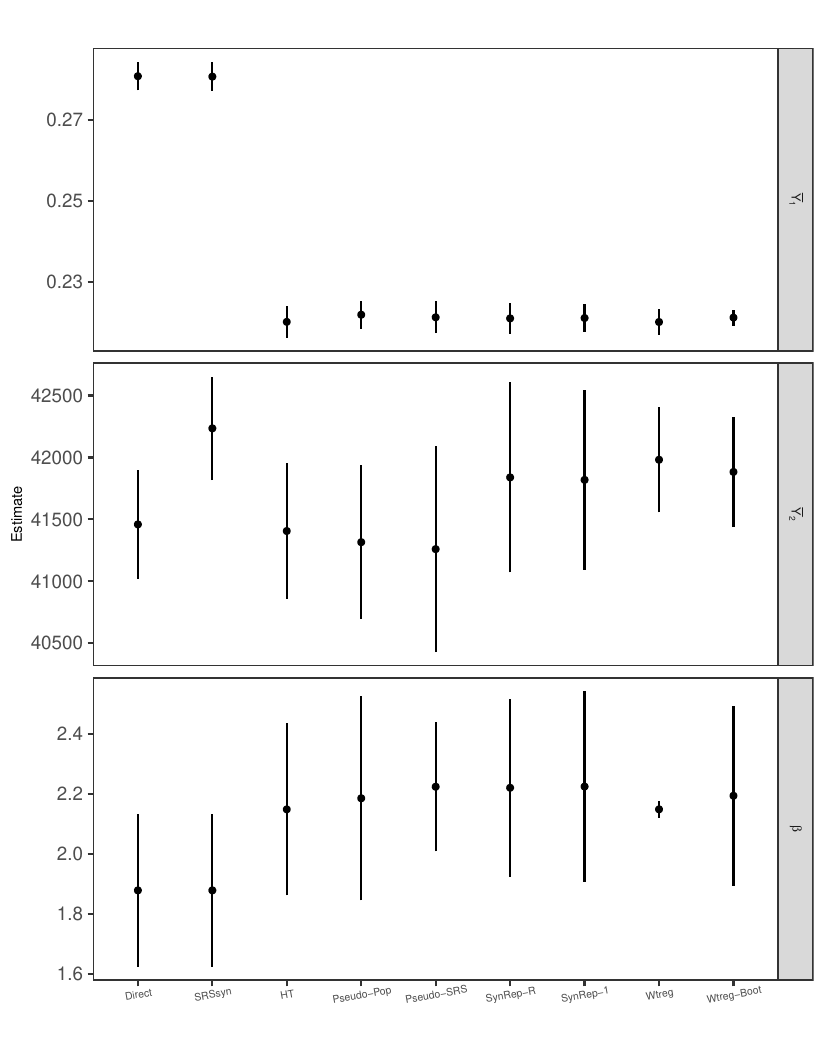}
\end{tabular}
\end{figure}

Figure~\ref{acs-m10r10} presents the point estimates and 95\% confidence intervals for the three population quantities. Since {\em Direct} and {\em SRSsyn} ignore the sample design, they result in relatively inaccurate results, especially for $\bar{Y}_1$. In contrast, the point estimates for the synthetic data methods that account for survey weights are closer to the {\em HT} point estimates. Additionally, the 95\% confidence intervals for these methods largely overlap with the {\em HT} confidence intervals.  We note, however, that {\em Wtreg} appears to suffer from underestimation of variance, particularly for $\beta$. Additionally, the confidence intervals for the pseudo-likelihood approaches can be narrower than those for {\em HT}, {\em SynRep-R}, and  {\em SynRep-1}.

\begin{table}[t]
\centering
\caption{Summaries of the differences (\$) in the largest income value in the synthetic and American Community Survey data.  The actual largest value is \$1,029,000.}
\label{risk}
\begin{tabular}{lrrrrrr}
  \hline
Method & Min. & 1st Quartile & Median & Mean & 3rd Quartile & Max. \\ 
  \hline
{\em SynRep-R} & -424,323  & -298,230  & -252,984  & -214,476  & -139,874  & 465,380  \\ 
{\em SynRep-1} & -371,466  & -297,180  & -287,199  & -268,711  & -267,428  & -40,405  \\ 
{\em Wtreg} & -440,253  & -297,689  & -242,095  & -218,810  & -159,766  & 707,411  \\ 
{\em Wtreg-Boot} & -410,354  & -275,398  & -209,513  & -174,444  & -139,109  & 133,759  \\ 
   \hline
\end{tabular}
\end{table}

We also can examine potential disclosure risks for the synthetic data methods. Here, we mimic an attack scenario described by \citet{SynInfor:Kim2021}, in which we consider an adversary who uses the synthetic data to estimate the largest income value in $\mathcal{D}$. Specifically, we examine differences between the maximum synthetic income in each synthetic dataset and the maximum income in $\mathcal{D}$. This evaluation is not intended to illustrate a rigorous and thorough process for assessing disclosure risks. Rather, we use this attack scenario mainly to compare the different synthesis procedures.

Table~\ref{risk} presents the distributions of the differences for the synthesis methods that account for the survey design. Overall, the results are reasonably similar across the methods, suggesting they offer similar levels of protection in this scenario.  All result in substantial differences between the largest synthetic and observed incomes. The results suggest that an adversary taking this attack strategy is not likely to estimate the largest income accurately.

\section{Discussion}
\label{discussion}

{\em SynRep-R} and {\em SynRep-1} represent a general strategy for constructing fully synthetic data that account for complex sample designs: use the WFPBB to ``undo'' the design, then replace the confidential values with simulated values. Releasing multiple synthetic data sets, i.e., setting $MR>1$, can increase statistical efficiency and facilitate variance estimation. However, agencies also can use the WFPBB approach with $MR=1$. Although releasing a single synthetic data set may not enable approximately valid variance estimation for complex surveys, it still can be useful in certain settings, e.g.,  when the synthetic data are intended for code training or exploratory analyses where variance estimation is not essential.

As noted by a reviewer, several agencies implementing synthetic data approaches also provide means for users to check the quality of their synthetic data inferences.  For example, users can submit their code to the agency that released the synthetic data, which then can run the code and report back disclosure-protected outputs to the user.  This is known as validation of results \citep{barrientos;etal;2017}.  Alternatively, users can submit queries to a server that computes an analysis of the confidential and synthetic data, and reports back measures of similarity of the two analysis results, e.g., the overlap in the confidence intervals \citep{utility}. This is known as verification of results \citep{barrientos;etal;2017}. With validation or verification, users of {\em SynRep-R} and {\em SynRep-1} may face an additional burden. If the agency directly runs the users' submitted analysis code, the user may need to specify a survey-weighted version of the code for validation, even though they have used a simple random sample analysis for synthetic data.  Of course, for many analyses, e.g., regression modeling, some users forego weighted analyses, in which case the issue is moot. It is also possible for the agency to automate validation or verification, in which case it may be able to turn users' submitted queries into survey-weighted versions automatically in the background; this is an area for future research.  

We chose to develop methods that enable agencies to follow the idea in \citet{rubin:1993}: release data that can be analyzed as simple random samples. This can make analyses easier for users, as they do not have to figure out how to deal with any weights on the file, e.g., in variance estimation. Releasing simple random samples could also help mitigate disclosure risks that may arise from releasing survey weights.  For example, if the weights released on the synthetic files 
are sampled directly from the weight values in $\mathcal{D}$ without alteration, the weights may reveal information about data subjects that is considered an unacceptable  disclosure risk \citep{Weightsforprivacy:Fienberg10}. Finally, releasing simple random samples avoids the need to estimate relationships between the weights and the outcome variables, which could be complicated in practice. Nonetheless, it would be interesting to compare risk and utility profiles of these approaches with those developed here.

There are many other topics related to the general strategy worth further investigation. First, in practice, survey weights can be highly variable and may not be strongly related to the survey variables of interest;  this can cause survey-weighted estimates to have inflated variances. This can be remedied somewhat, for example, by using model-based approaches to smooth the weights~\citep{beaumont08,elliot:JOS16,prior-si2018}. Synthetic data generation based on the WFPBB (or any other approach) is not immune to these weighting issues. Thus, it would be interesting to examine if and how the synthesis model can reduce the effects of variance inflation from extreme weights. 

Second, we focus on developing the fully synthetic data framework and corresponding combining rules, using simple settings and synthesis models to illustrate the methods. Conceptually, agencies can apply {\em SynRep-R} and {\em SynRep-1} to multivariate data and for various estimands of interest, e.g., subdomain means and multiple regression coefficients. In such cases, it may be advantageous to use flexible modeling approaches, such as tree-based models or other machine learning algorithms.  Future work could investigate the performance of these synthesizers in combination with the pseudo-population and pseudo-SRS generation steps. 

Third, we derive the combining rules assuming the original survey data are complete. Agencies could impute missing survey data and generate synthetic replicates simultaneously, possibly accounting for the complex design in the imputation model and synthesis approach.  This strategy may necessitate new combining rules akin to those in \citet{reitermi}.  

Fourth, we present {\em ad hoc} adjustments to deal with negative values of the variance estimates.  We may be able to improve on those adjustments. For example, we may be able to adapt the strategy in \citet{sireiter11}, who develop inferential methods for fully synthetic data based on sampling from the distributions used in the derivations of the combining rules. Additionally, as pointed out by a reviewer,  it may be beneficial to use the insight of \cite{raab2016practical} for the sampling and synthesis components of the derivation in {\em SynRep-R}.  This results in an alternative variance estimator, $\left(1+M^{-1} \right)b_{syn} - \left(1+R^{-1} \right)\bar{v}_{syn}.$ Future work can investigate the performance of these alternative inference methods.

Fifth, it would be informative to generalize the implementation of {\em SynRep-R} and {\em Syn-Rep-1} to other complex designs, such as the stratified multi-stage cluster sampling designs that are common in practice. \cite{Zhou:JOS16} have extended the WFPBB to account for strata, clustering, and survey weights in synthetic population generation. We expect that one could take simple random samples from these pseudo-populations and generate synthetic replicates, possibly using synthesis models that capture design information as suggested in \cite{reiter:2002}, and extend the combining rules presented here. It would be a natural extension to comprehensively assess the repeated sampling performances of {\em SynRep-R} and {\em Syn-Rep-1} in such multi-stage complex samples. 

Lastly, it would be useful to develop principled approaches to measuring disclosure risks for these methods.  For {\em SynRep-R} and {\em SynRep-1}, conceptually one could estimate an adversary's posterior distribution for confidential data values given the released synthetic values, e.g., as described for simple settings in \citet{synriskzhang} and \citet{hureiter14}. However, this would be computationally challenging in practice. One would need to account for the entire synthetic data generation process---including the bootstrapping, sampling, and synthesis---when computing this posterior distribution.  Indeed, as far as we are aware, agencies that release synthetic data use {\em ad hoc} approaches to assessing disclosure risks, such as comparing the similarity of outlier values in the confidential and synthetic data as we illustrated here \citep{kinney:reiter:miranda}.  Developing disclosure risk methods is a major area for future research for all approaches to generating fully synthetic data.

\section*{Acknowledgements}

The work was funded by the U.S. National Science Foundation grant (SES 2217456) and a pilot project from the Michigan Center on the Demography of Aging with funding from the National Institute on Aging (P30 AG012846).

\appendix

\section{Appendix: Additional Simulation Results}

\begin{figure}[htp]
\caption{Standard deviation (SD) of estimated (est) variances of different population quantities with different procedures for different numbers of synthetic samples ($M$) and replicates ($R$) under a PPS design.}
\label{pps-m10r10-sd}
\begin{tabular}{c}
\includegraphics[width=0.9\textwidth]{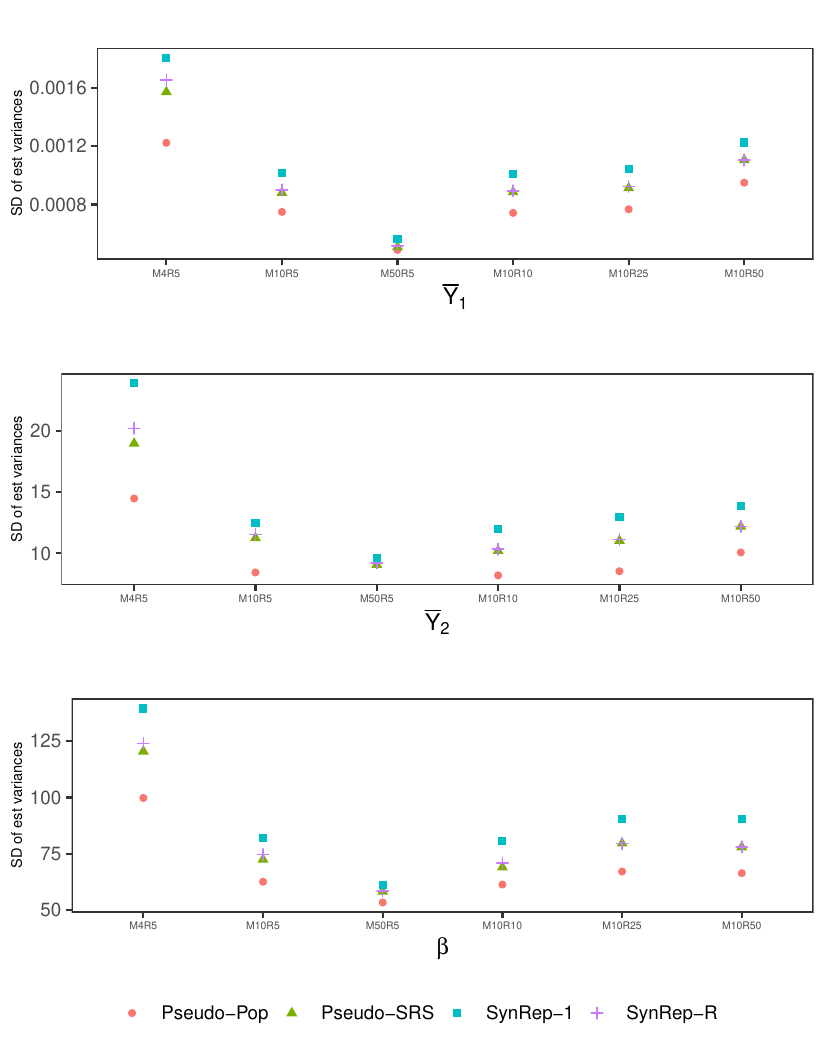}
\end{tabular}
\end{figure}

Figure~\ref{pps-m10r10-sd} displays the variability of the 1000 values of estimated variances of the point estimators for $\beta, \bar{Y}_1$, and $\bar{Y}_2$ for the simulation with the PPS design. The variability tends to decrease with $M$.  Increasing $R$ when $M$ is held constant seems not to have much impact on the stability of the results. We see increased variability as the procedures introduce more steps that involve randomness; that is, as we go from {\em Pseudo-Pop} to {\em Pseudo-SRS} to {\em SynRep-R} and {\em SynRep-1}. The variability tends to be largest for {\em SynRep-1}.

\begin{figure}[htp]
\caption{Repeated sampling properties of different quantities and procedures with $M=10$ synthetic samples and $R=10$ replicates under a SRS design. }
\label{srs-m10r10}
\begin{tabular}{c}
\includegraphics[width=0.95\textwidth]{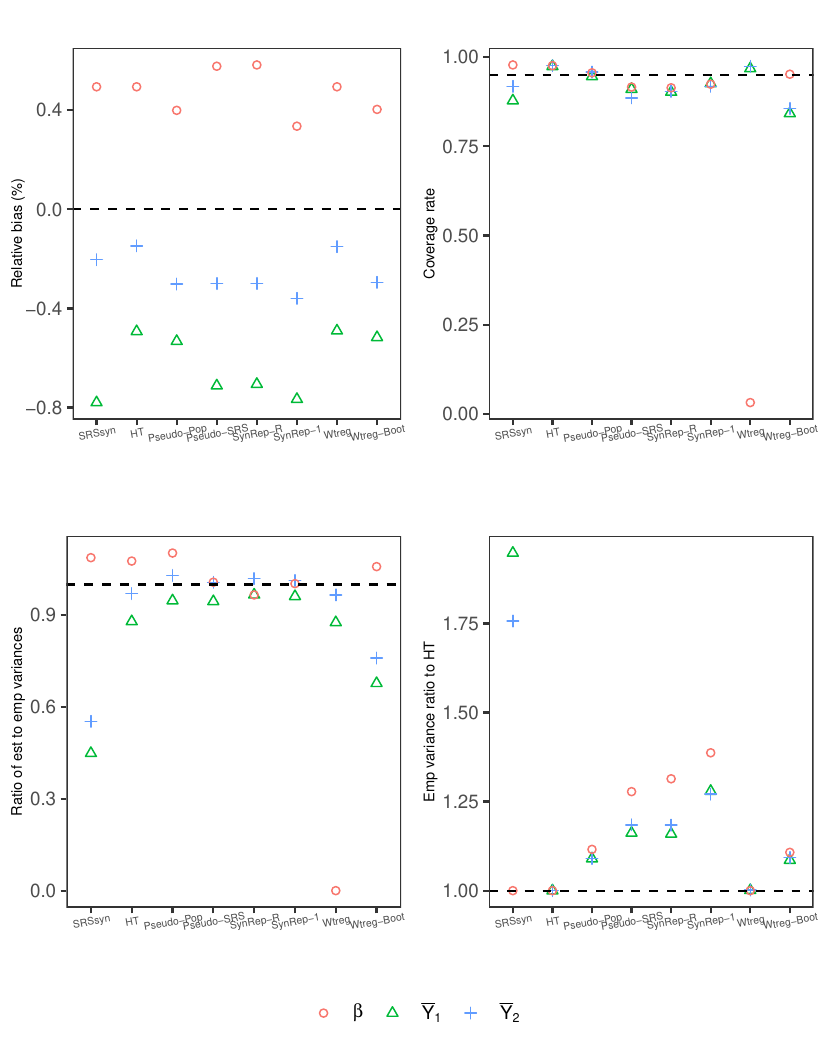}
\end{tabular}
\end{figure}

As another check of the validity of the combining rules, we repeat the simulations from Section \ref{simulation} using a SRS in place of a PPS design. Specifically, we use the population described in Section \ref{sim:design}, but we use a SRS of $n=500$ records for each $\mathcal{D}$. 
Figure~\ref{srs-m10r10} displays the results. Overall, the performances of {\em SynRep-R} and {\em SynRep-1} mirror the patterns seen for the PPS design in Section \ref{simulation}.

\bibliography{ys-2023June-b.bib}
\bibliographystyle{chicago}







\end{document}